\documentclass[aps,prd,twocolumn,groupedaddress]{revtex4}
\usepackage{hyperref}
\usepackage{graphicx}% Include figure files
\usepackage{dcolumn}% Align table columns on decimal point
\usepackage{subfigure}
\usepackage{bm}% bold math

\begin{document}

\pacs{Valid PACS appear here}
\title{CMB Tomography: Reconstruction of Adiabatic Primordial Scalar Potential Using Temperature and Polarization Maps}
\author{A. P. S. Yadav}%
\affiliation{Department of Astronomy, University of Illinois at Urbana-Champaign, 1002 W.~Green Street, Urbana, IL 61801}

\author{B. D. Wandelt}
\affiliation{Department of Astronomy, University of Illinois at Urbana-Champaign, 1002 W.~Green Street, Urbana, IL 61801}
\affiliation{Department of Physics,  University of Illinois at Urbana-Champaign, 1110 W.~Green Street, Urbana, IL 61801}
\affiliation{Center of Advanced Studies, 912, W.~Illinois Street, Urbana, IL 61801}

\begin{abstract}
Assuming linearity of the perturbations at the time of decoupling, we reconstruct the primordial scalar potential from the temperature and polarization anisotropies in the cosmic 
microwave background radiation. In doing so we derive an optimal linear filter
which, when operated on the spherical harmonic coefficients of the anisotropy
maps, returns an estimate of the primordial scalar potential
fluctuations in a spherical slice. The reconstruction is best in a thick shell around the decoupling epoch; the quality of the
reconstruction depends on the redshift of the slice within this shell. With
high quality maps of the temperature and polarization anisotropies it will be
possible to obtain a reconstruction of potential fluctuation on scales between
$\ell=2$ and $\ell \sim 300$ at the redshift
of decoupling, with some information about the three-dimensional shapes of the perturbations in
a shell of width 250Mpc. 

\end{abstract}
\maketitle
\section{Introduction}
Cosmic microwave background (CMB) radiation has given us a great deal of insight into the early universe \cite{cmb}, for 
example 
the pattern of 
temperature 
fluctuation gives us the information about the perturbations in the plasma at
a temperature $T \sim 3000$K. Using the 
concordance between big bang nucleosynthesis and CMB \cite{bbn1}, we are able to probe
directly the
earlier universe $T \sim 10^{10}$K, but we rely on an understanding of
cosmological theory to learn about the Universe at even higher energy
scales. One of the most promising theories of the early universe is inflation  
\cite{Guth81}, which predicts
(i) spatial flatness of the observable universe today, (ii) nearly scale
invariant, adiabatic, primordial density perturbations, (iii) homogeneity and isotropy  on large angular scales of the
observable universe, and (iv) Gaussianity.
Predictions (i) and (ii) have been  
observationally tested \cite{flatscale}, while predictions (iii) and (iv) are
yet to be conclusively tested.

Any characterization of the Gaussianity (or otherwise) of the primordial perturbations will constrain models of 
inflation (e.g. \cite{inf_gauss1,inf_gauss2,inf_gauss3}). Any non-Gaussianity predicted in canonical
inflation is very small \cite{Acquaviva02,Maldacena02}, but inflation
scenarios with large amounts of non-Gaussianity can be constructed
\cite{nong1,nong2,nong3}. So far, bispectrum
tests for primordial non-Gaussianity have not detected a signal in the
temperature fluctuations mapped by COBE \cite{nong_bdw} and WMAP 
\cite{nong_wmap}. Other authors have found non-Gaussianity signatures in the
WMAP temperature data \cite{hotcold,larsonwandelt,nong_cmb1,nong_cmb2,nong_cmb3}. 
The main motivation of reconstructing primordial 
potential perturbations is to study these 
non-Gaussianities. Reconstruction allows 
us to be more sensitive to the primordial
perturbations, which is important because current detections of non-Gaussianity
do not specifically select for the primordial perturbations. An advantage of
our approach is the ability to use the reconstructed maps to study higher order 
statistics of the primordial perturbations directly in the reconstructed
potential.

We generalize the results from \cite{4} to use both the temperature and
polarization information. We will show that this 
results in a 3D reconstruction of a thick slice centered at $z_{dec}$. Using the polarization information removes
the blind spots which are present at certain scales in the temperature-only
analysis. We provide examples that show what will be possible with a future
experiment that provide high-fidelity maps of the temperature
and E-polarization anisotropies. 

We assume
that the CMB  two-point 
statistics (i.e. power spectrum) and other sources of cosmological information will reveal the parameters to a high
degree of accuracy such that they are input parameters in our reconstruction. If desired our technique can  be generalized to
include residual uncertainties in cosmological parameters by coupling it to
Markov chains.

\section{Reconstruction of Primordial Fluctuation using Temperature and Polarization}
The harmonic coefficients of the CMB anisotropy $(a_{lm})$ are related to 
primordial fluctuation as:
\begin{equation} 
\label{1}
a_{\ell m}^X=\frac{2}{\pi}\int k^2dk\,r^2dr\,\lbrack \, \Phi_{\ell m}(r) \, g^{adi}_{X\ell}(k) + S_{\ell m}(r) 
\,g^{iso}_{X\ell}(k) \, 
\rbrack \, j_\ell(kr)\end{equation}
where $\Phi_{\ell m}(r)$ and  $S_{\ell m}(r) $ are, respectively, the curvature and the isocurvature harmonic 
coefficients of the potential at a given comoving distance 
$r=\vert \mathbf {r} \vert$; $g_{X\ell}(r)$ is the radiation 
transfer function of either adiabatic or isocurvature perturbations; and X refers to either T or E. \\ \\
Assuming that curvature perturbations $\Phi(\mathbf {\vec{r}})$ dominate over isocurvature perturbations $S(\mathbf 
{\vec{r}})$, 
we try to reconstruct the scalar potential $\Phi(r)$ 
from the observed temperature and polarization of the CMB. A  linear filter $O_{\ell}(r)$ which when operated on 
$a_{\ell m}$ 
reconstructs the primordial field,  can be obtained 
by minimizing the difference between the filtered field $O_{\ell}(r)a_{\ell m}$ and the underlying field 
$\Phi_{\ell m}(r)$ \cite{4}. 
Since 
the 
perturbations were in the linear regime when the CMB decoupled, the choice of
a linear filter is justified. Positive definiteness 
of $\frac {\partial ^{2}}{\partial {O_{\ell}^X(r)}^{2}}$ guarantees a minima of chi square at each $\ell$ and hence 
guarantees 
the 
existence of $O_\ell^{X}$. 

\begin{equation} \frac {\partial}{\partial O_{\ell}^X(r)} \langle \arrowvert \sum_{X=T,E} \, 
O_{\ell}^X(r)\,a_{\ell m}^X-\Phi_{\ell m}(r) 
\arrowvert^2 
\rangle =0 \label{2}\end{equation}
which gives following form for the filters:

\begin{equation} \left( \begin{array}{c} O_{\ell}^T(r)\\O_{\ell}^E(r)\\ \end{array} \right) = \left( \begin{array}{ccc} 
C_{\ell}^{TT}&C_{\ell}^{TE}\\C_{l}^{TE}&C_{l}^{EE}\\\end{array}
\right)^{-1}\, \left( \begin{array}{c} \beta_{\ell}^T(r)\\ \beta_{\ell}^E(r) \\ \end{array} \right), \label{3}\end{equation}
where
\begin{equation} C_{\ell}^{XY}=\langle  a_{\ell m}^X a_{\ell m}^{*Y}\rangle=\frac {2}{\pi} \int k^2 dk \,P_{\Phi}(k) 
\,g_{X\ell}\,g_{Y\ell}(k) 
\end{equation}
\begin{equation} \beta_{\ell}^X(r)=\langle  \Phi_{\ell m}(r) a_{\ell m}^{*X}\rangle=\frac {2}{\pi} \int k^2 dk 
\,P_{\Phi}(k) \,g_{X\ell}(k)\, 
j_{\ell}(kr), \end{equation}
$j_{\ell}(kr)$ being the Bessel function of order $\ell$ and $P_{\Phi}(k)$ is the primordial power spectrum of scalar 
metric 
perturbations. In the presence of the instrumental noise, the expression for  $C^{XY}_{\ell}$ modifies. Assuming that the noise is 
Gaussian 
with 
variance 
$\bar{\sigma}^{X}_{\ell m}$, and spatially homogenious; we can write $\bar{\sigma}^{X}_{\ell m} = 
\bar{\sigma}^{X}_{0}\delta_{\ell \ell ^{\prime}}\delta_{mm^{\prime}}$ then $C^{XX}_{\ell}$ is replaced by $C^{XX}_{\ell} + 
(\bar{\sigma}^{X}_{0})^{2}$. We have assumed that $\bar{\sigma}^{TE}_{lm}$  is zero.      

This defines an estimator for the potential
\begin{equation} \hat{\Phi}_{\ell m}(r) = \sum_{X=T,E}{O_{\ell}^X(r)}\,a_{\ell m}^{X}\label{6} \end{equation}

As evident the equations in (3) are coupled linear equations with variables $ O_{\ell}^{T}(r) $ and $ O_{\ell}^{E}(r) $.\\ \\
If only one of the available CMB maps (either temperature T or polarization E) is used to reconstruct the 
potential, 
these equations decouple into ordinary linear 
equations in one variable $ O_{\ell}^{X}(r) $ where X can either be T or E. In these special cases Eq. \ref{2} modifies to 
\begin{equation} \frac 
{\partial}{\partial O_{\ell}^X(r)} \langle \arrowvert  \, O_{\ell}^X(r)\,a_{\ell m}^X-\Phi_{\ell m}(r)
\arrowvert^2
\rangle =0\end{equation}and gives the following form for the filter:

\begin{figure}
\includegraphics[height=0.5\textheight, width=0.5\textwidth, angle=0]{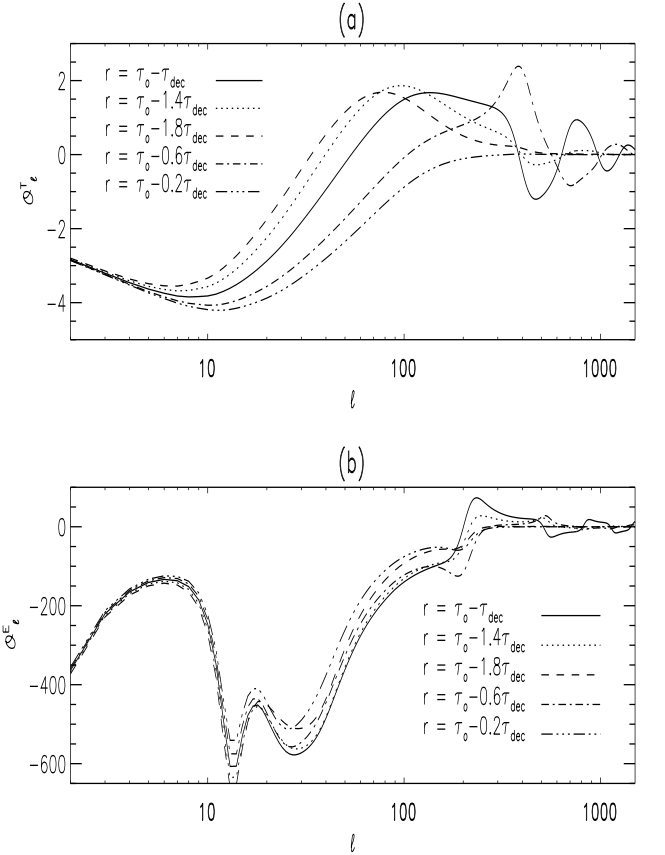}
\caption{Filters using T and E separately, at five different conformal
distances, $r= c(\tau_o -\tau_{dec})$, where $\tau_o$ is the present conformal
time and $\tau_{dec}$ is the conformal time at photon decoupling epoch. We use
a $\Lambda$CDM model with re-ionization and 
$c\tau_o=14.2827$Gpc and $c\tau_{dec}=0.2836$Gpc.
(a)$O_{\ell}^T(r)=\frac{\beta_{\ell}^T(r)}{C_{\ell}^{TT}}$
(b)$O_{\ell}^E(r)=\frac {\beta_{\ell}^E(r)}{C_{\ell}^{EE}}$ \label{fig1}}
\end{figure}

\begin{equation} O_{\ell}^X(r)=\frac {\beta_{\ell}^X(r)}{C_{\ell}^{XX}} \label{8}\end{equation}

We have first reconstructed the metric perturbations using the temperature and polarization information separately as 
given by Eq. \ref{8}. Then we solved Eq. \ref {6} to combine the information
from temperature T and 
E polarization and show that reconstruction is much better. 

Using both T and E, the  expected variance of the reconstructed potential is given by

\begin{equation} \sigma^2_{lm}(r)= \Big\{ \sum_X \frac{(O_{\ell}^X(r))^2}{\bar{\sigma}^{X}_{\ell m}} + 
(P^{\phi}_{\ell}(r))^{-1} 
\Big\}^{-1},  
\label{9} 
\end{equation}
where $\bar{\sigma}^T_{\ell m}$ and $\bar{\sigma}^E_{\ell m}$ are the  noise variances in the temperature map and 
in the polarization map respectively. We define 
\begin{equation}
\hat{P}^{\Phi}_{\ell}= \frac {1}{2\ell +1} \sum^{\ell}_{m=-\ell} \hat{\Phi}_{\ell m} \hat{\Phi} ^{\ast}_{\ell m}
\end{equation}and use its expectation value:
\begin{equation}
\left < \hat{P}^{\Phi}_{\ell} \right >= (O_{\ell}^T(r))^{2}C_{\ell}^{TT} 
+(O_{\ell}^E(r))^{2}C_{\ell}^{EE}+2O_{\ell}^T(r)O_{\ell}^{E}(r)C_{\ell}^{TE},
\end{equation}for the purpose of comparing our reconstruction to the theoretical power spectrum. 
As an estimator of the primordial power spectrum this biased because our
filters are optimized for estimating the potential 
$\Phi_{\ell m}(r)$  and not the power 
spectrum. This is the standard for optimal, or Wiener, filters, since they
remove contamination aggressively. One can compute the bias to correct for
this effect and bound the allowed range for the true power spectrum as 
$\left \langle \hat{P}^{\phi}_{\ell} \right \rangle \pm 
\sigma^{2}_{\ell}$, where
\begin{eqnarray}
\sigma^{2}_{\ell}(r)= \sqrt {2}\Big\{ P^{\phi}_{\ell}(r) + (O_{\ell}^T(r))^{2}C_{\ell}^{TT} +
(O_{\ell}^E(r))^{2}C_{\ell}^{EE}\nonumber \\
-2\Big(
O_{\ell}^T(r)\beta_{\ell}^T(r)+O_{\ell}^E(r)\beta_{\ell}^E(r)-O_{\ell}^T(r)O_{\ell}^E(r)\beta_{\ell}^{TE}(r) \Big) \Big\}.
\label{10}
\end{eqnarray}
Here, $ P^{\phi}_{\ell}(r)$ is the theory power
spectrum of the potential on a spherical shell of radius $r$.

\begin{figure*}[t]
\includegraphics[height=73mm]{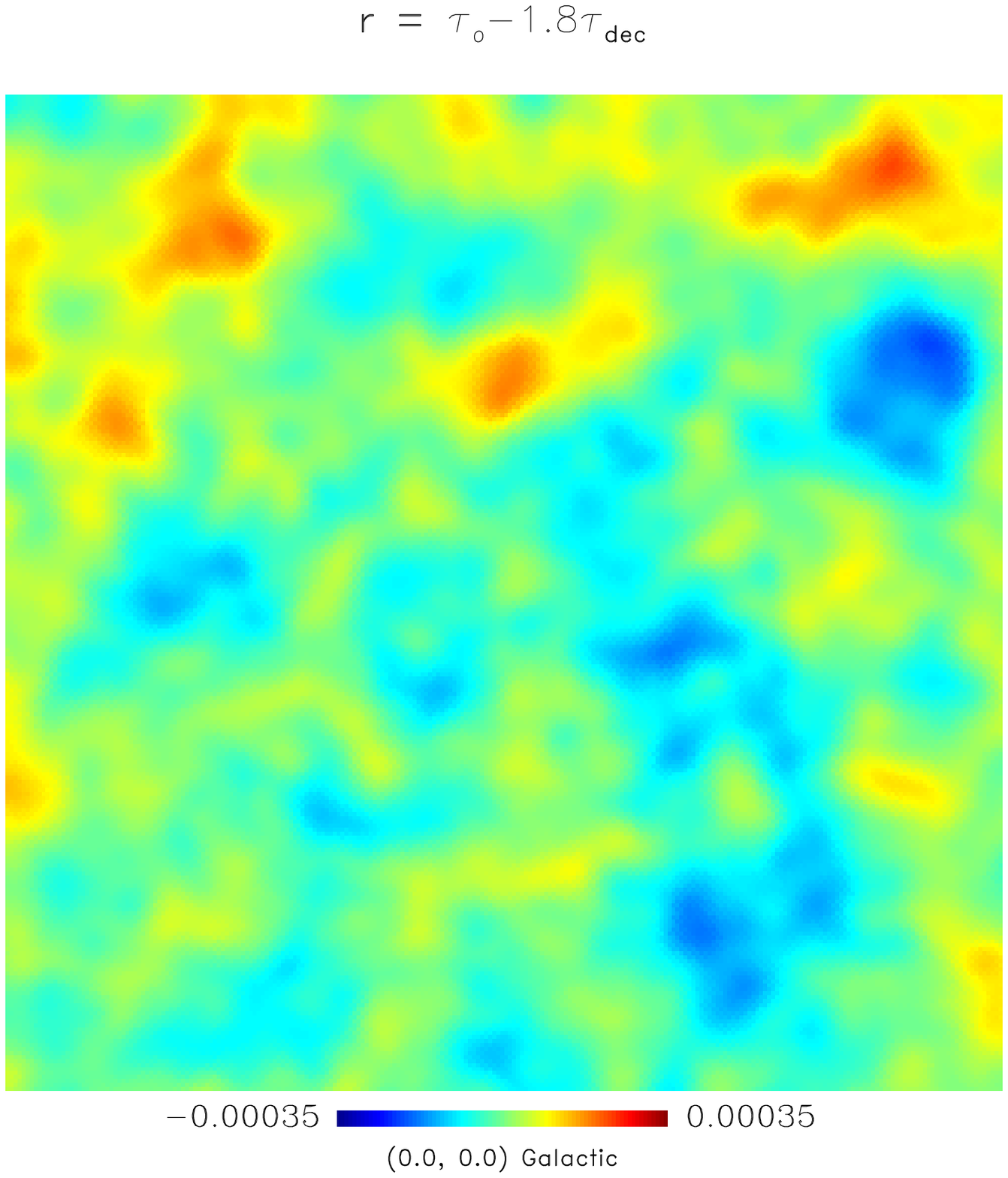} 
\includegraphics[height=73mm]{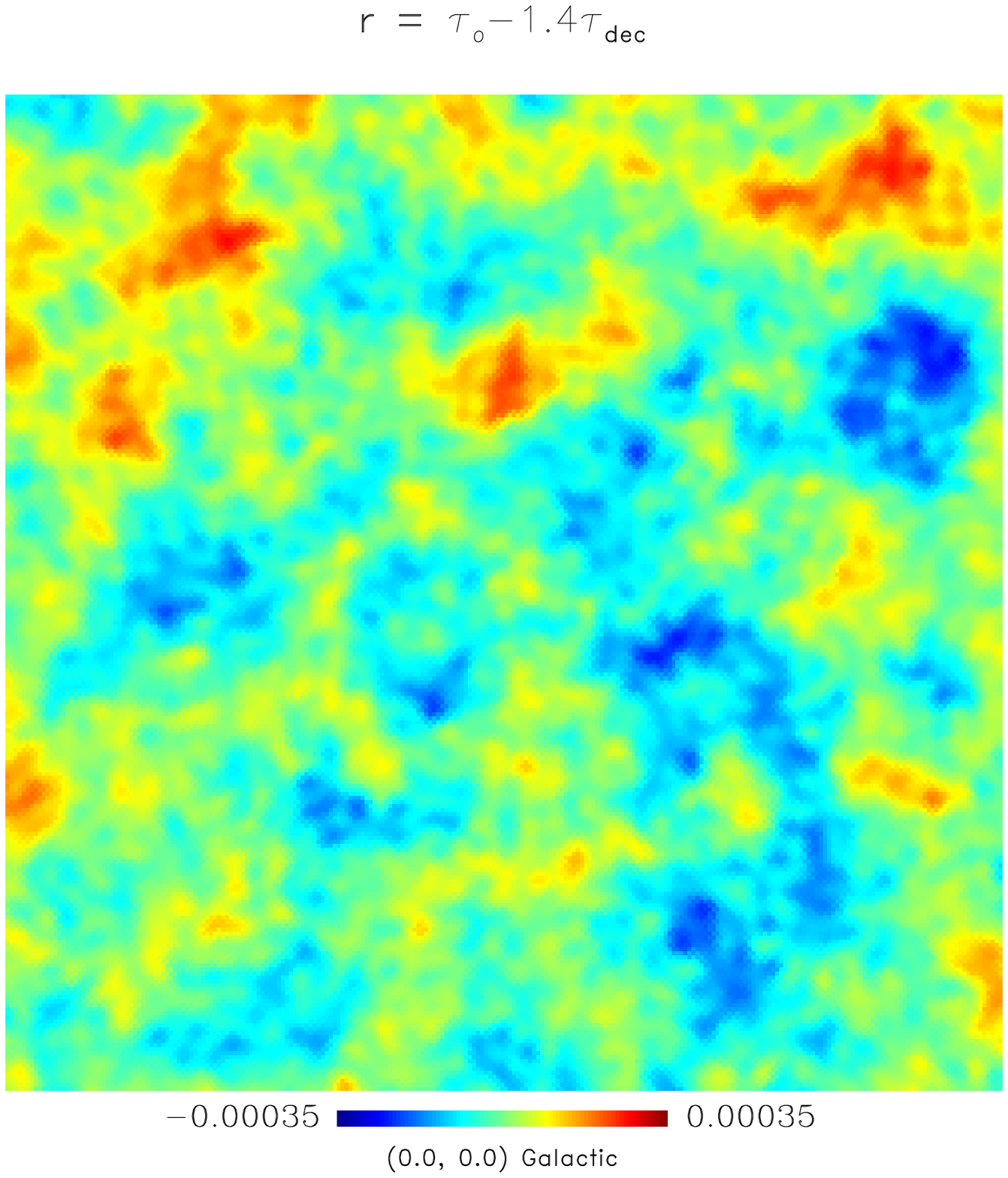}
\includegraphics[height=73mm]{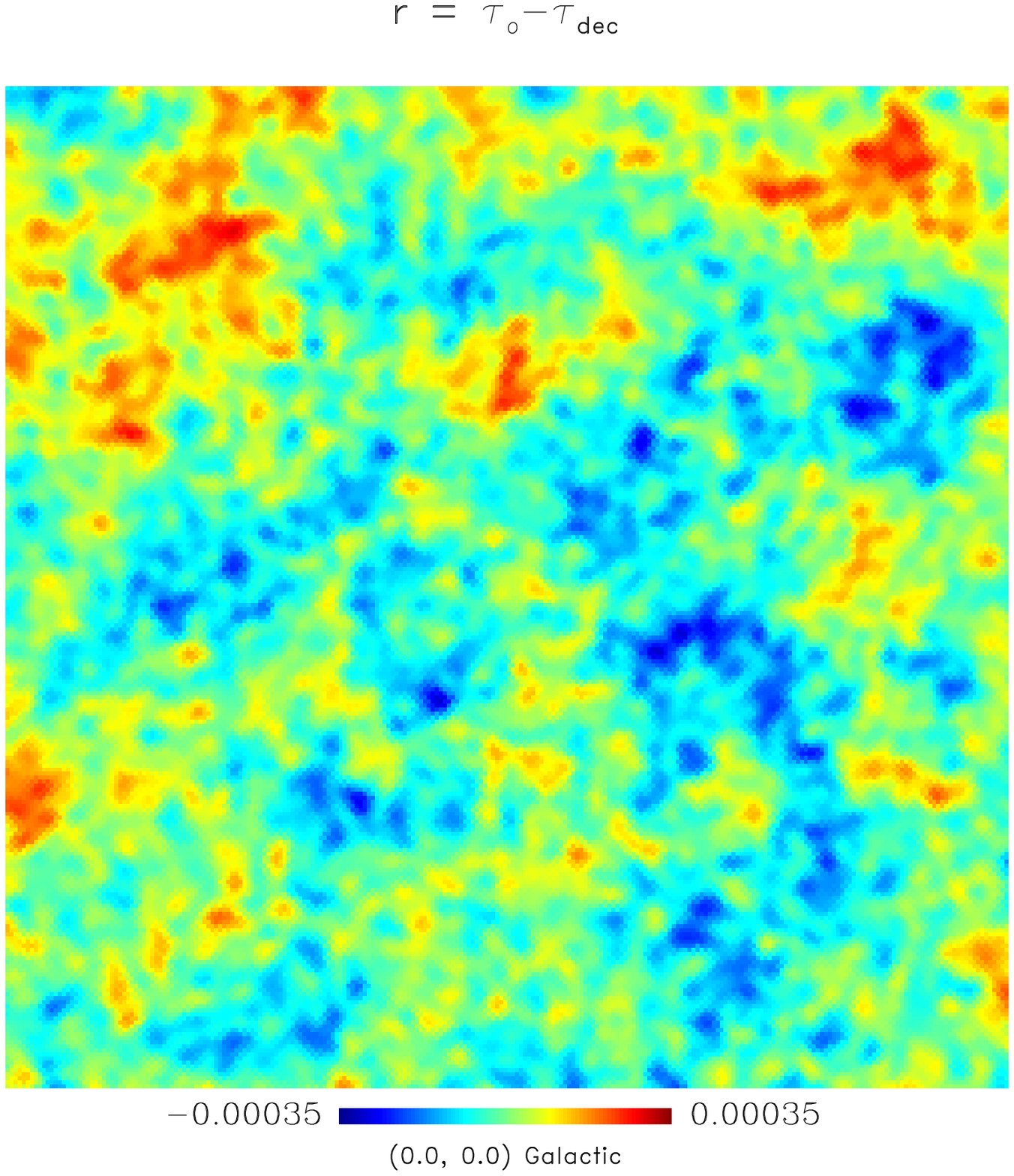}
%\end{figure}
%\begin{figure}[t]
\includegraphics[height=73mm]{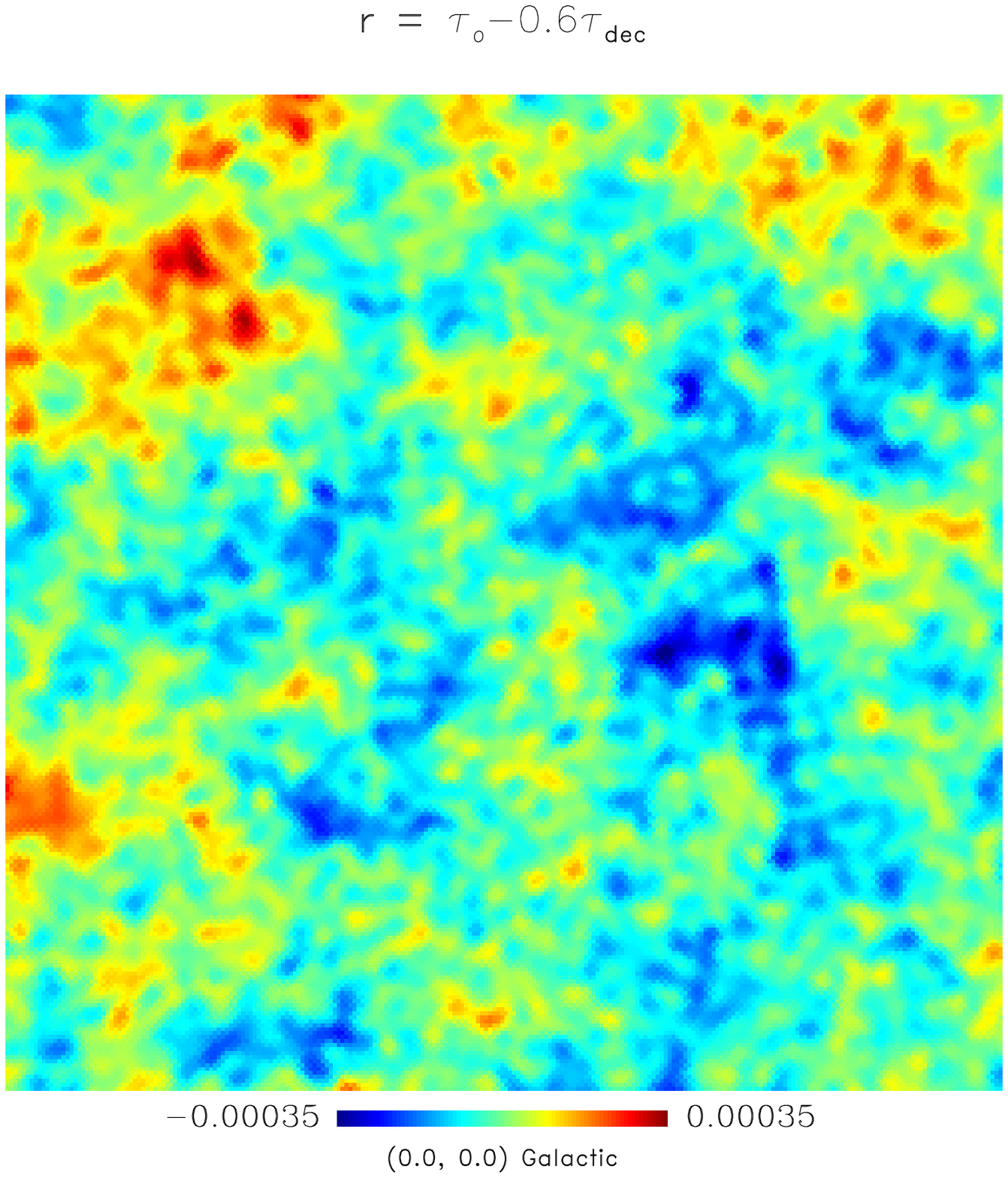}
\includegraphics[height=73mm]{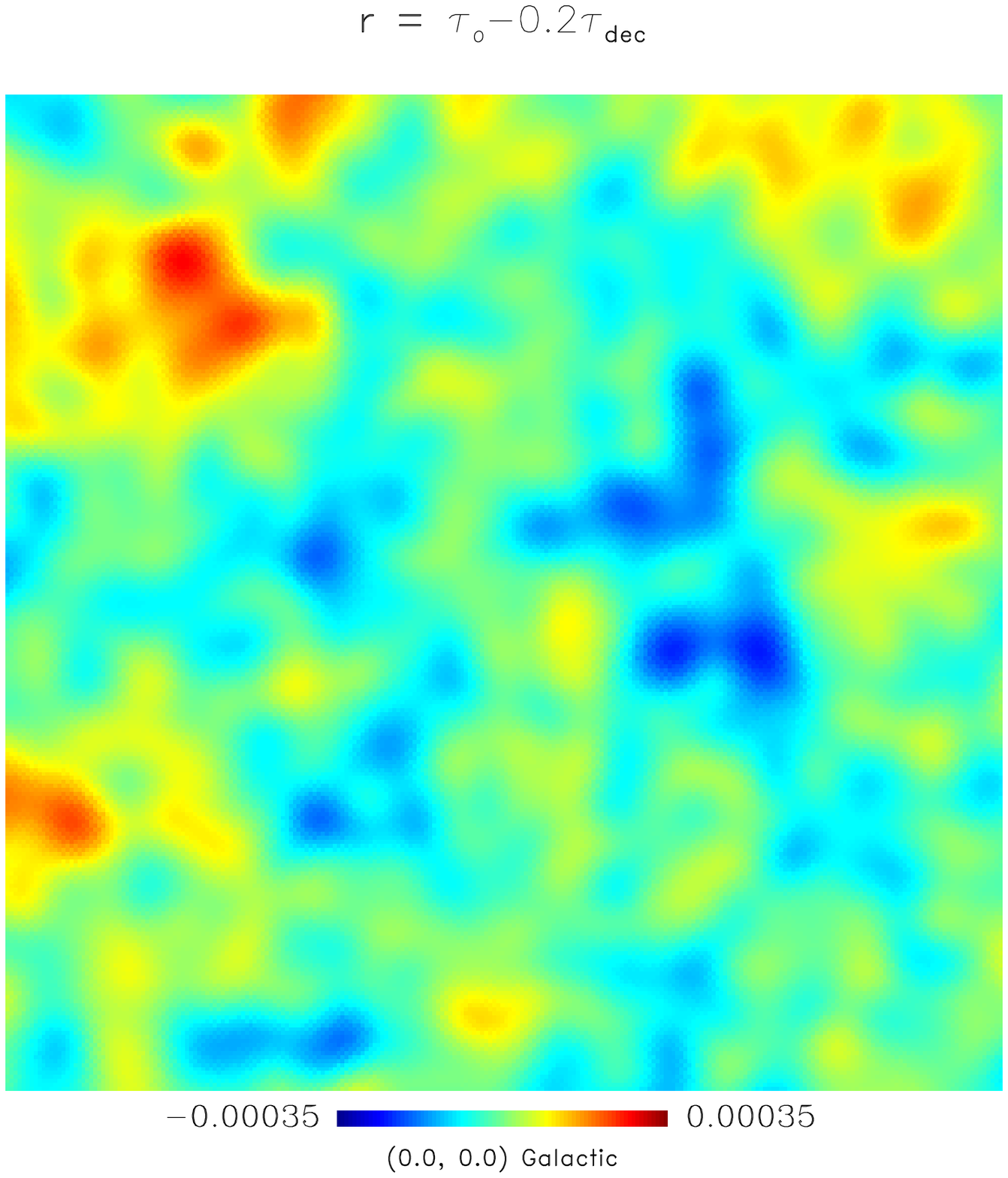}
%\caption{Same as Fig. 2}
\caption{Reconstructed potential maps in slices centered around the sphere of last scattering
$r_{dec}= c(\tau_o -\tau_{dec})$. The maps are ordered in decreasing distance,
left-to-right and top-to-bottom. The color scale is in dimensionless units of fractional perturbation. Each map is 25 degrees on the side.}
\end{figure*}

\section{Results}
Our examples are computed within  
a $\Lambda $CDM
model with and without re-ionization ($\Omega_c=0.26, \Omega_b=0.04, \Omega_{\Lambda}=0.7, h=0.7$, scalar spectral index 
$n_s=0.93$, with running 
$\frac{dn_s}{dlnk}=-0.031$, re-ionization optical depth $\tau=0.17$).

Figure 1 shows plots of the operators $O_{\ell}^X(r)$ (for X as T and E respectively) for different  r as a function of 
$\ell$. 
While the
reconstruction of the primordial potential will be good on large scales for both T and E (agreeing 
with the Sachs-Wolfe effect \cite{6}) , it will be bad on small scales as $O_{\ell}^X(r)$ oscillates about 0 for both T 
and E. But 
as a consequence of the physics at decoupling, the transfer functions $g_{T\, \ell}(k)$ and $g_{E\,
  \ell}(k)$ are out of phase. This results in 
$O_{\ell}^T(r)$ and $ O_{\ell}^E(r)$  never both being zero at a particular
scale (see Figures 3f and 4f).
This observation supports our analysis leading to the coupled Eq. \ref{3} and we are  
able to reconstruct  the potential on much broader range of scales if we use both temperature and
polarization 
information from CMB. 

In Figure 2  we show reconstructed potential maps at various slices of
conformal time. The quality of reconstruction is sensitive to the
reconstruction depth and is best in the 
thin slice centered on the peak of the differential optical depth. By stacking
these slices a three-dimensional thick shell of primordial potential is
reconstructed.

\begin{figure*}
\includegraphics[height=.72\textheight,width=.865\textwidth]{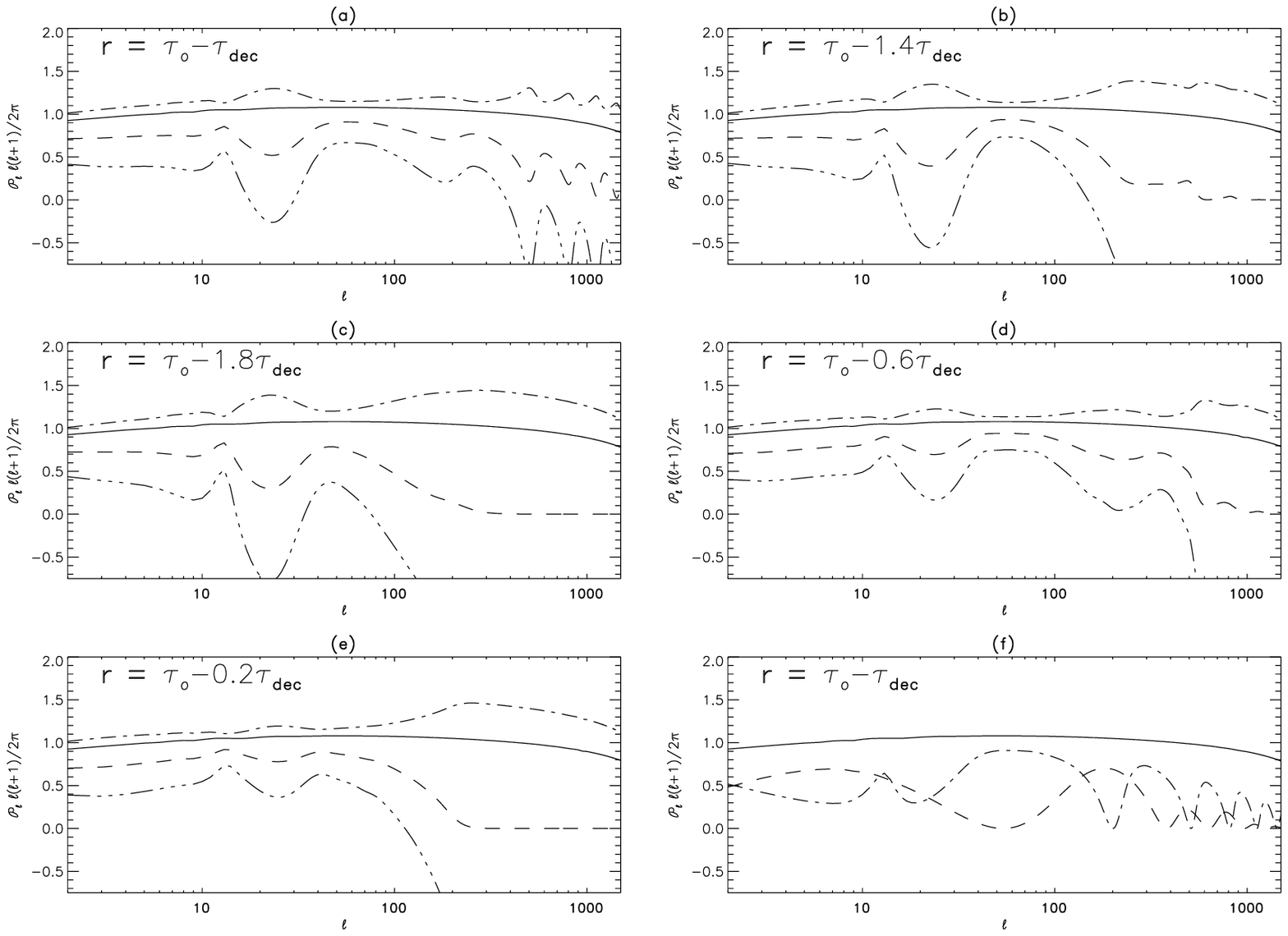}
\caption{Power spectrum of the primordial fluctuation for the $\Lambda $CDM
model with re-ionization ($\Omega_c=0.26, \Omega_b=0.04, \Omega_{\Lambda}=0.7,
h=0.7$, $\tau = 0.17$, $n_s=0.93$, $\frac{dn_s}{dlnk}=-0.031$). For all plots
the solid line shows the theoretical primordial power spectrum. In (a)-(e) we
show the reconstructed power spectrum using the coupled estimator given by
Eq. (\ref{6}) (dashed), bracketed by error bounds (dot-dashed). In (f) we show
the reconstructions due to temperature (dashed) and polarization (dot-dashed)
alone, using the estimator given by Eq. (\ref{8}).  The theory and the
reconstructed power spectrum are shown in arbitrary units.
\label{fig2}}
\end{figure*}

\begin{figure*}
\includegraphics[height=.72\textheight,width=0.865\textwidth]{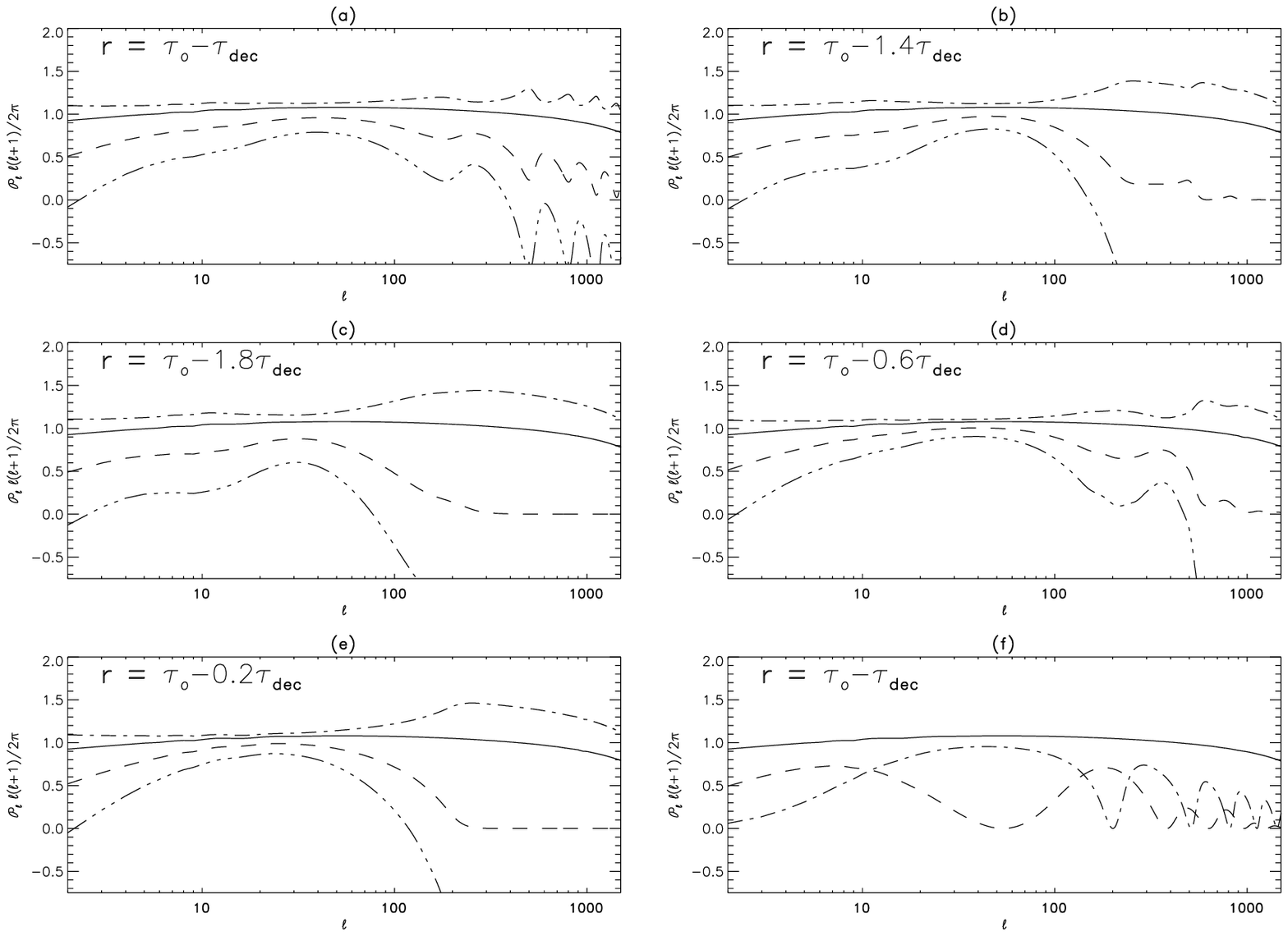}
\title{aaa}
\caption{Power spectrum of the primordial fluctuation for $\Lambda $CDM model
without re-ionization ($\Omega_c=0.26, \Omega_b=0.04, \Omega_{\Lambda}=0.7,
h=0.7$, $\tau = 0$,$n_s=0.93$, $\frac{dn_s}{dlnk}=-0.031$).  For all plots the
solid line shows the theoretical primordial power spectrum. In (a)-(e) we show
the reconstructed power spectrum using the coupled estimator given by
Eq. (\ref{6}) (dashed), bracketed by error bounds (dot-dashed). In (f) we show
the reconstructions due to temperature (dashed) and polarization (dot-dashed)
alone, using the estimator given by Eq. (\ref{8}).  The theory and the
reconstructed power spectrum are shown in arbitrary units.
\label{fig3}}
\end{figure*}

Figures 3 and 4 show the expected power spectra of the reconstructed potential
maps for models with and without re-ionization. The power spectra are bounded from above by the theoretical power
spectrum computed from our cosmological parameters $n_s$ and $A$. The amount
of reconstructed power is indicative of the quality of the
reconstruction. As discussed in section II, our filter does not minimize the error bars on this power spectrum. The fact that
the error bars include the true spectrum show that the reconstruction of the perturbation itself is unbiased.
At small angular scales $\ell\sim 300$ subhorizon physics wash out the information about the primordial
fluctuations and the reconstruction fails.

The combination of temperature and E polarization data enables
the reconstruction up to $\ell\sim 300$, without blind spots in $\ell$. This is a marked
improvement compared to the performance of the operator that uses either
temperature or polarization
alone which are shown in Figures 3f and 4f.

In Figures 3 and 4 the error bars increase due to cosmic variance on large angular scales. In Figure 3
we see that re-ionization improves the ability to
reconstruct the  primordial potential at large scales (except in a trough near
$\ell \sim 20$). The polarization-only reconstruction reveals that
the enlarged E polarization signal due to early re-ionization aids in the
reconstruction at very low $\ell$. Physically, the rescattering of photons at low redshift imprints information on the
large scale modes of the primordial perturbation \cite{Skordis04} .

%%\section{Discussion}
%%In our analysis we have assumed that there exist a linear operator which can construct the primordial perturbations. The 
%%motivation for such an operator is
%%(i) that the perturbations were linear at the time of decoupling and (ii) after 
%%decoupling, CMB fluctuations were not modified (except through expansion
%%effects).
%%
%%We assume that integral given by 
%%Eq. (\ref{1}) has the most dominant contribution in a small $z$ centered 
%%around 
%%$z_{dec}$. But the CMB did not travel unhindered after decoupling, and effects like re-ionization and 
%%gravitational lensing make our assumption an approximation. If we know about
%%re-ionization then our operator would combine  giving a contribution around $z_{dec}$, we will have more
%%operators contributing also at $z_{reion}$. Since we 
%%yet do not know much about the re-ionization mechanism, we can turn this problem around and ask following question: what new
%%operator (in addition to the one which we derived) is required to reconstruct the primordial power spectrum (NOT 
%%perturbations). Then the new operator should have the information about the
%%details of the re-ionization history.

\section{Conclusion and Future Work}
We have demonstrated the application of  linear theory to construct a least-square
estimator of the primordial potential of adiabatic density fluctuations
generated during the epoch of inflation. Starting with high-quality maps of
the CMB anisotropies in temperature and E polarization and using a fiducial set of cosmological
parameters we can reconstruct the primordial potential in a spherical  shell of
thickness $\sim$ 250Mpc near  the peak of the differential optical depth. Reconstruction is reasonably good up to $\ell \sim
300$ and the stronger E polarization signal due to re-ionization aids in
reconstructing the potential at very low $\ell$. This effect is consistent with
and a generalization of
the results by Skordis and Silk \cite{Skordis04} who
reconstruct the CMB quadrupole with
accuracy better than cosmic variance.
%%To
%%improve 
%%the quality of reconstruction we need information about the re-ionization and gravitational lensing 
%%effects on 

To exploit the tomographic techniques we have developed, we require maps of T
and E and high signal-to-noise ratios. For this first exploration we have 
neglected the effect of foregrounds, which will reduce the effective area of
the T and E maps that we will be able to use in the reconstruction. One of the
advantages of CMB analysis techniques based on Gibbs sampling
\cite{jewell04,wll04} is that methods such as this one can be applied
essentially without modification while the messy details of dealing with
systematics like  foregrounds and instrumental contaminants
are taken into account by the Bayesian analysis.

One might wonder whether there are ways to improve on our techniques. In
principle this could be done in two distinct ways. First, we might wish to
broaden the range in redshift over which we can reconstruct the primordial
perturbations. A second way to improve on our techniques would be to enhance
the quality of reconstruction. 

On the first issue, scattering of cosmic microwave
background photons through processes other than Thompson scattering, such as
Rayleigh scattering or resonant scattering makes the location of the last scattering surface frequency
dependent\cite{SpergelOstriker01,SunyaevEtAl05}. In principle, this type of effect would
allow a more detailed tomography of the inflationary perturbations over a
broader range in redshift. However, in the models currently favored by
cosmological observations Rayleigh scattering only varies the redshift of the
peak in the visibility function by about 1\% \cite{SpergelOstriker01}. This is
much less than the thickness of the last scattering surface, and hence does
not broaden the redshift range where our reconstruction is effective. A
detailed study of the promise of using resonant scattering for this purpose remains
to be done.

On the second issue, improving the quality of the reconstruction, it is possible that additional
linear combinations of primordial perturbation modes
could be constrained by high signal-to-noise
CMB maps over a range of frequencies.

Ultimately we envision tests of non-Gaussianity applied to the reconstructed
primordial potential. Our filters will select those combinations of the
spherical harmonic modes that correlate to the primordial potential. This
reduces the
probability that residual foregrounds result in a non-Gaussian
signature. Further, the ability to reconstruct the potential in a thick shell
allows the application of three dimensional shape statistics to potential peaks and
troughs. We anticipate the search for non-Gaussianities in the  primordial perturbations once high resolution and high sensitivity maps
of the E polarization signal become available. To our knowledge the techniques we describe in this paper represent
the most direct way to date to probe
the statistical properties of the primordial
perturbations created in the inflation era.

\acknowledgments
We acknowledge conversations in the early stages of this project with
M.~Sazhin and R.~Caldwell. We acknowledge the use of the CMBFAST package by  Uros Seljak and Matias Zaldarriaga \cite{5}. Some of the 
results in 
this paper have been derived using the HEALPix (G\'orski, Hivon, and Wandelt 1999) package. 

%%\bibliography{references}

\begin{thebibliography}{27}
\expandafter\ifx\csname natexlab\endcsname\relax\def\natexlab#1{#1}\fi
\expandafter\ifx\csname bibnamefont\endcsname\relax
  \def\bibnamefont#1{#1}\fi
\expandafter\ifx\csname bibfnamefont\endcsname\relax
  \def\bibfnamefont#1{#1}\fi
\expandafter\ifx\csname citenamefont\endcsname\relax
  \def\citenamefont#1{#1}\fi
\expandafter\ifx\csname url\endcsname\relax
  \def\url#1{\texttt{#1}}\fi
\expandafter\ifx\csname urlprefix\endcsname\relax\def\urlprefix{URL }\fi
\providecommand{\bibinfo}[2]{#2}
\providecommand{\eprint}[2][]{\url{#2}}

\bibitem[{\citenamefont{Bennett et~al.}(2003)\citenamefont{Bennett, Halpern,
  Hinshaw, Jarosik, Kogut, Limon, Meyer, Page, Spergel, Tucker et~al.}}]{cmb}
\bibinfo{author}{\bibfnamefont{C.~L.} \bibnamefont{Bennett}},
  \bibinfo{author}{\bibfnamefont{M.}~\bibnamefont{Halpern}},
  \bibinfo{author}{\bibfnamefont{G.}~\bibnamefont{Hinshaw}},
  \bibinfo{author}{\bibfnamefont{N.}~\bibnamefont{Jarosik}},
  \bibinfo{author}{\bibfnamefont{A.}~\bibnamefont{Kogut}},
  \bibinfo{author}{\bibfnamefont{M.}~\bibnamefont{Limon}},
  \bibinfo{author}{\bibfnamefont{S.~S.} \bibnamefont{Meyer}},
  \bibinfo{author}{\bibfnamefont{L.}~\bibnamefont{Page}},
  \bibinfo{author}{\bibfnamefont{D.~N.} \bibnamefont{Spergel}},
  \bibinfo{author}{\bibfnamefont{G.~S.} \bibnamefont{Tucker}},
  \bibnamefont{et~al.}, \bibinfo{journal}{Astrophys. J. S.}
  \textbf{\bibinfo{volume}{148}}, \bibinfo{pages}{1} (\bibinfo{year}{2003}).

\bibitem[{\citenamefont{Cyburt et~al.}(2003)\citenamefont{Cyburt, Fields, and
  Olive}}]{bbn1}
\bibinfo{author}{\bibfnamefont{R.~H.} \bibnamefont{Cyburt}},
  \bibinfo{author}{\bibfnamefont{B.~D.} \bibnamefont{Fields}},
  \bibnamefont{and} \bibinfo{author}{\bibfnamefont{K.~A.} \bibnamefont{Olive}},
  \bibinfo{journal}{Phys. Lett. B} \textbf{\bibinfo{volume}{567}},
  \bibinfo{pages}{227} (\bibinfo{year}{2003}).

\bibitem[{\citenamefont{Guth}(1981)}]{Guth81}
\bibinfo{author}{\bibfnamefont{A.~H.} \bibnamefont{Guth}},
  \bibinfo{journal}{Phys. Rev. D} \textbf{\bibinfo{volume}{23}},
  \bibinfo{pages}{347} (\bibinfo{year}{1981}).

\bibitem[{\citenamefont{Spergel et~al.}(2003)\citenamefont{Spergel, Verde,
  Peiris, Komatsu, Nolta, Bennett, Halpern, Hinshaw, Jarosik, Kogut
  et~al.}}]{flatscale}
\bibinfo{author}{\bibfnamefont{D.~N.} \bibnamefont{Spergel}},
  \bibinfo{author}{\bibfnamefont{L.}~\bibnamefont{Verde}},
  \bibinfo{author}{\bibfnamefont{H.~V.} \bibnamefont{Peiris}},
  \bibinfo{author}{\bibfnamefont{E.}~\bibnamefont{Komatsu}},
  \bibinfo{author}{\bibfnamefont{M.~R.} \bibnamefont{Nolta}},
  \bibinfo{author}{\bibfnamefont{C.~L.} \bibnamefont{Bennett}},
  \bibinfo{author}{\bibfnamefont{M.}~\bibnamefont{Halpern}},
  \bibinfo{author}{\bibfnamefont{G.}~\bibnamefont{Hinshaw}},
  \bibinfo{author}{\bibfnamefont{N.}~\bibnamefont{Jarosik}},
  \bibinfo{author}{\bibfnamefont{A.}~\bibnamefont{Kogut}},
  \bibnamefont{et~al.}, \bibinfo{journal}{Astrophys. J. S.}
  \textbf{\bibinfo{volume}{148}}, \bibinfo{pages}{175} (\bibinfo{year}{2003}).

\bibitem[{\citenamefont{Starobinsky}(1982)}]{inf_gauss1}
\bibinfo{author}{\bibfnamefont{A.~A.} \bibnamefont{Starobinsky}},
  \bibinfo{journal}{Phys. Lett. B} \textbf{\bibinfo{volume}{117}},
  \bibinfo{pages}{175} (\bibinfo{year}{1982}).

\bibitem[{\citenamefont{Guth and Pi}(1982)}]{inf_gauss2}
\bibinfo{author}{\bibfnamefont{A.~H.} \bibnamefont{Guth}} \bibnamefont{and}
  \bibinfo{author}{\bibfnamefont{S.~Y.} \bibnamefont{Pi}},
  \bibinfo{journal}{Phys. Rev. Lett.} \textbf{\bibinfo{volume}{49}},
  \bibinfo{pages}{1110} (\bibinfo{year}{1982}).

\bibitem[{\citenamefont{Bardeen et~al.}(1983)\citenamefont{Bardeen, Steinhardt,
  and Turner}}]{inf_gauss3}
\bibinfo{author}{\bibfnamefont{J.~M.} \bibnamefont{Bardeen}},
  \bibinfo{author}{\bibfnamefont{P.~J.} \bibnamefont{Steinhardt}},
  \bibnamefont{and} \bibinfo{author}{\bibfnamefont{M.~S.}
  \bibnamefont{Turner}}, \bibinfo{journal}{Phys. Rev. D.}
  \textbf{\bibinfo{volume}{28}}, \bibinfo{pages}{679} (\bibinfo{year}{1983}).

\bibitem[{\citenamefont{Acquaviva et~al.}(2002)\citenamefont{Acquaviva,
  Botrolo, Matarrese, and Riotto}}]{Acquaviva02}
\bibinfo{author}{\bibfnamefont{V.}~\bibnamefont{Acquaviva}},
  \bibinfo{author}{\bibfnamefont{N.}~\bibnamefont{Botrolo}},
  \bibinfo{author}{\bibfnamefont{S.}~\bibnamefont{Matarrese}},
  \bibnamefont{and} \bibinfo{author}{\bibfnamefont{A.}~\bibnamefont{Riotto}}
  (\bibinfo{year}{2002}), \eprint{astro-ph/0209156}.

\bibitem[{\citenamefont{Maldacena}(2002)}]{Maldacena02}
\bibinfo{author}{\bibfnamefont{J.}~\bibnamefont{Maldacena}}
  (\bibinfo{year}{2002}), \eprint{astro-ph/0210603}.

\bibitem[{\citenamefont{Salopek and Bond}(1990)}]{nong1}
\bibinfo{author}{\bibfnamefont{D.~S.} \bibnamefont{Salopek}} \bibnamefont{and}
  \bibinfo{author}{\bibfnamefont{J.~R.} \bibnamefont{Bond}},
  \bibinfo{journal}{Phys. Rev. D} \textbf{\bibinfo{volume}{42}},
  \bibinfo{pages}{3936} (\bibinfo{year}{1990}).

\bibitem[{\citenamefont{Gangui et~al.}(1994)\citenamefont{Gangui, Lucchin,
  Matarrese, and Mollerach}}]{nong2}
\bibinfo{author}{\bibfnamefont{A.}~\bibnamefont{Gangui}},
  \bibinfo{author}{\bibfnamefont{F.}~\bibnamefont{Lucchin}},
  \bibinfo{author}{\bibfnamefont{S.}~\bibnamefont{Matarrese}},
  \bibnamefont{and}
  \bibinfo{author}{\bibfnamefont{S.}~\bibnamefont{Mollerach}},
  \bibinfo{journal}{Astrophys. J.} \textbf{\bibinfo{volume}{430}},
  \bibinfo{pages}{447} (\bibinfo{year}{1994}).

\bibitem[{\citenamefont{Gangui}(1994)}]{nong3}
\bibinfo{author}{\bibfnamefont{A.}~\bibnamefont{Gangui}},
  \bibinfo{journal}{Phys. Rev. D} \textbf{\bibinfo{volume}{50}},
  \bibinfo{pages}{3684} (\bibinfo{year}{1994}).

\bibitem[{\citenamefont{Komatsu et~al.}(2002)\citenamefont{Komatsu, Wandelt,
  Spergel, Banday, and Gorski}}]{nong_bdw}
\bibinfo{author}{\bibfnamefont{E.~N.} \bibnamefont{Komatsu}},
  \bibinfo{author}{\bibfnamefont{B.~D.} \bibnamefont{Wandelt}},
  \bibinfo{author}{\bibfnamefont{D.~N.} \bibnamefont{Spergel}},
  \bibinfo{author}{\bibfnamefont{A.~J.} \bibnamefont{Banday}},
  \bibnamefont{and} \bibinfo{author}{\bibfnamefont{K.~M.}
  \bibnamefont{Gorski}}, \bibinfo{journal}{Astrophys. J.}
  \textbf{\bibinfo{volume}{566}}, \bibinfo{pages}{19} (\bibinfo{year}{2002}).

\bibitem[{\citenamefont{Komatsu
  et~al.}(2003{\natexlab{a}})\citenamefont{Komatsu, Kogut, Nolta, Bennett,
  Halpern, Hinshaw, Jarosik, Limon, Meyer, Page et~al.}}]{nong_wmap}
\bibinfo{author}{\bibfnamefont{E.~N.} \bibnamefont{Komatsu}},
  \bibinfo{author}{\bibfnamefont{A.}~\bibnamefont{Kogut}},
  \bibinfo{author}{\bibfnamefont{M.}~\bibnamefont{Nolta}},
  \bibinfo{author}{\bibfnamefont{C.~L.} \bibnamefont{Bennett}},
  \bibinfo{author}{\bibfnamefont{M.}~\bibnamefont{Halpern}},
  \bibinfo{author}{\bibfnamefont{G.}~\bibnamefont{Hinshaw}},
  \bibinfo{author}{\bibfnamefont{N.}~\bibnamefont{Jarosik}},
  \bibinfo{author}{\bibfnamefont{M.}~\bibnamefont{Limon}},
  \bibinfo{author}{\bibfnamefont{S.~S.} \bibnamefont{Meyer}},
  \bibinfo{author}{\bibfnamefont{L.}~\bibnamefont{Page}}, \bibnamefont{et~al.},
  \bibinfo{journal}{Astrophys. J.} \textbf{\bibinfo{volume}{148}},
  \bibinfo{pages}{119} (\bibinfo{year}{2003}{\natexlab{a}}).

\bibitem[{\citenamefont{Mukherjee and Wang}(2004)}]{hotcold}
\bibinfo{author}{\bibfnamefont{P.}~\bibnamefont{Mukherjee}} \bibnamefont{and}
  \bibinfo{author}{\bibfnamefont{Y.}~\bibnamefont{Wang}},
  \bibinfo{journal}{Astrophys. J.} \textbf{\bibinfo{volume}{613}},
  \bibinfo{pages}{51} (\bibinfo{year}{2004}).

\bibitem[{\citenamefont{Larson and Wandelt}(2004)}]{larsonwandelt}
\bibinfo{author}{\bibfnamefont{D.~L.} \bibnamefont{Larson}} \bibnamefont{and}
  \bibinfo{author}{\bibfnamefont{B.~D.} \bibnamefont{Wandelt}},
  \bibinfo{journal}{Astrophys. J.} \textbf{\bibinfo{volume}{613}},
  \bibinfo{pages}{L85} (\bibinfo{year}{2004}).

\bibitem[{\citenamefont{Vielva et~al.}(2004)\citenamefont{Vielva, Gonzalez,
  Barreiro, Sanz, and Cayon}}]{nong_cmb1}
\bibinfo{author}{\bibfnamefont{P.}~\bibnamefont{Vielva}},
  \bibinfo{author}{\bibfnamefont{E.~M.} \bibnamefont{Gonzalez}},
  \bibinfo{author}{\bibfnamefont{R.~B.} \bibnamefont{Barreiro}},
  \bibinfo{author}{\bibfnamefont{J.~L.} \bibnamefont{Sanz}}, \bibnamefont{and}
  \bibinfo{author}{\bibfnamefont{L.}~\bibnamefont{Cayon}},
  \bibinfo{journal}{Astrophys. J} \textbf{\bibinfo{volume}{609}},
  \bibinfo{pages}{22} (\bibinfo{year}{2004}).

\bibitem[{\citenamefont{Chiang et~al.}(2003)\citenamefont{Chiang, P.~D,
  Verkhodanov, and Way}}]{nong_cmb2}
\bibinfo{author}{\bibfnamefont{L.~Y.} \bibnamefont{Chiang}},
  \bibinfo{author}{\bibfnamefont{N.}~\bibnamefont{P.~D}},
  \bibinfo{author}{\bibfnamefont{O.~V.} \bibnamefont{Verkhodanov}},
  \bibnamefont{and} \bibinfo{author}{\bibfnamefont{M.~J.} \bibnamefont{Way}},
  \bibinfo{journal}{Astrophys. J} \textbf{\bibinfo{volume}{590}},
  \bibinfo{pages}{65} (\bibinfo{year}{2003}).

\bibitem[{\citenamefont{Chiang et~al.}(2004)\citenamefont{Chiang, Naselsky, and
  Coles}}]{nong_cmb3}
\bibinfo{author}{\bibfnamefont{L.~Y.} \bibnamefont{Chiang}},
  \bibinfo{author}{\bibfnamefont{P.~D.} \bibnamefont{Naselsky}},
  \bibnamefont{and} \bibinfo{author}{\bibfnamefont{P.}~\bibnamefont{Coles}},
  \bibinfo{journal}{Astrophys. J} \textbf{\bibinfo{volume}{602}},
  \bibinfo{pages}{1} (\bibinfo{year}{2004}).

\bibitem[{\citenamefont{Komatsu
  et~al.}(2003{\natexlab{b}})\citenamefont{Komatsu, Spergel, and Wandelt}}]{4}
\bibinfo{author}{\bibfnamefont{E.~N.} \bibnamefont{Komatsu}},
  \bibinfo{author}{\bibfnamefont{D.~N.} \bibnamefont{Spergel}},
  \bibnamefont{and} \bibinfo{author}{\bibfnamefont{B.~D.}
  \bibnamefont{Wandelt}} (\bibinfo{year}{2003}{\natexlab{b}}),
  \eprint{astro-ph/0305189}.

\bibitem[{\citenamefont{Sachs and Wolfe}(1967)}]{6}
\bibinfo{author}{\bibfnamefont{R.~K.} \bibnamefont{Sachs}} \bibnamefont{and}
  \bibinfo{author}{\bibfnamefont{A.~M.} \bibnamefont{Wolfe}},
  \bibinfo{journal}{Astrophys. J} \textbf{\bibinfo{volume}{147}},
  \bibinfo{pages}{73} (\bibinfo{year}{1967}).

\bibitem[{\citenamefont{Skordis and Silk}(2004)}]{Skordis04}
\bibinfo{author}{\bibfnamefont{C.}~\bibnamefont{Skordis}} \bibnamefont{and}
  \bibinfo{author}{\bibfnamefont{J.}~\bibnamefont{Silk}}
  (\bibinfo{year}{2004}), \eprint{astro-ph/0402474}.

\bibitem[{\citenamefont{Jewell et~al.}(2004)\citenamefont{Jewell, Levin, and
  Anderson}}]{jewell04}
\bibinfo{author}{\bibfnamefont{J.}~\bibnamefont{Jewell}},
  \bibinfo{author}{\bibfnamefont{S.}~\bibnamefont{Levin}}, \bibnamefont{and}
  \bibinfo{author}{\bibfnamefont{C.~H.} \bibnamefont{Anderson}},
  \bibinfo{journal}{Astrophys. J.} \textbf{\bibinfo{volume}{609}},
  \bibinfo{pages}{1} (\bibinfo{year}{2004}).

\bibitem[{\citenamefont{Wandelt et~al.}(2004)\citenamefont{Wandelt, Larson, and
  Lakshminarayanan}}]{wll04}
\bibinfo{author}{\bibfnamefont{B.~D.} \bibnamefont{Wandelt}},
  \bibinfo{author}{\bibfnamefont{D.~L.} \bibnamefont{Larson}},
  \bibnamefont{and}
  \bibinfo{author}{\bibfnamefont{A.}~\bibnamefont{Lakshminarayanan}},
  \bibinfo{journal}{Phys. Rev. D.} \textbf{\bibinfo{volume}{70}},
  \bibinfo{pages}{083511} (\bibinfo{year}{2004}).

\bibitem[{\citenamefont{Yu et~al.}(2001)\citenamefont{Yu, Spergel, and
  Ostriker}}]{SpergelOstriker01}
\bibinfo{author}{\bibfnamefont{Q.}~\bibnamefont{Yu}},
  \bibinfo{author}{\bibfnamefont{D.~N.} \bibnamefont{Spergel}},
  \bibnamefont{and} \bibinfo{author}{\bibfnamefont{J.~P.}
  \bibnamefont{Ostriker}}, \bibinfo{journal}{Astrophys. J.}
  \textbf{\bibinfo{volume}{558}}, \bibinfo{pages}{23} (\bibinfo{year}{2001}).

\bibitem[{\citenamefont{Hernández-Monteagudo and
  Sunyaev}(2005)}]{SunyaevEtAl05}
\bibinfo{author}{\bibfnamefont{C.}~\bibnamefont{Hernández-Monteagudo}}
  \bibnamefont{and} \bibinfo{author}{\bibfnamefont{R.~A.}
  \bibnamefont{Sunyaev}}, \bibinfo{journal}{MNRAS}
  \textbf{\bibinfo{volume}{359}}, \bibinfo{pages}{597} (\bibinfo{year}{2005}).

\bibitem[{\citenamefont{Seljak and Zaldarriaga}(1996)}]{5}
\bibinfo{author}{\bibfnamefont{U.}~\bibnamefont{Seljak}} \bibnamefont{and}
  \bibinfo{author}{\bibfnamefont{M.}~\bibnamefont{Zaldarriaga}},
  \bibinfo{journal}{Astrophys. J} \textbf{\bibinfo{volume}{469}},
  \bibinfo{pages}{437} (\bibinfo{year}{1996}).

\end{thebibliography}

\end{document}